\documentclass[CEJP,PDF]{cej} 
\usepackage{layout}
\usepackage{amsmath}
\usepackage{textcomp}

\title{Nonextensive perfect hydrodynamics
 - a model of dissipative relativistic hydrodynamics?}

\articletype{Research Article}

\author{Takeshi Osada\inst{1}\email{osada@ph.ns.musashi-tech.ac.jp},
        Grzegorz Wilk\inst{2}\email{wilk@fuw.edu.pl}}

\institute{
     \inst{1} Theoretical Physics Lab., Faculty of Knowledge Engineering,
Musashi Institute of Technology,\\
     Setagaya-ku, Tokyo 158-8557, Japan
     \inst{2} The Andrzej So\l tan Institute for Nuclear Studies,
     Theoretical Physics Department,\\
     ul. Ho\.za 69, 00-681 Warsaw, Poland
          }

\abstract{We demonstrate that nonextensive perfect relativistic
hydrodynamics ($q$-hydrodynamics) can serve as a model of the
usual relativistic dissipative hydrodynamics ($d$-hydrodynamics)
facilitating therefore considerably its applications. As
illustration we show how using $q$-hydrodynamics one gets the
$q$-dependent expressions for the dissipative entropy current and
the corresponding ratios of the bulk and shear viscosities to
entropy density, $\zeta/s$ and $\eta/s$.}


\begin{document}
\maketitle


\section{Introduction}

Hydrodynamics is well a established effective approach to flow
phenomena. Here we shall present its nonextensive relativistic
version from the point of view of high energy collision physics
\cite{Landau,CW,recenthydro,HB,TC,hydro-review}. The
characteristic feature of such a processes is the production of a
large number of secondaries (their multiplicities at present
approach $\sim 10^3$). Already in 1953, when there were only $\sim
10$ particles produced and registered, mostly in cosmic rays
experiments, it was found \cite{Landau} that such processes can be
very effectively described by using a thermodynamic approach in
conjunction with hydrodynamical flow. In this so called Landau
Hydrodynamic Model the secondaries secondaries were considered to
be a product of decay of some kind of hadronic fluid produced in
such collisions, which was expanding before hadronization
\cite{Landau}. This model has recently been updated to describe
recent experimental data \cite{CW}. Since then there was a number
of successful attempts to develop new solutions for both the
Landau model \cite{recenthydro} and for the so called Hwa-Bjorken
version of the hydrodynamical model \cite{HB}, for which a new
class of solutions has been found \cite{TC}. Hydrodynamic models
of different types were therefore frequently used in the
phenomenological description of multiparticle production processes
at high energies, especially for high energy nuclear collisions
\cite{hydro-review}. These are of special interest to us (being
currently investigated at RHIC at Brookhaven with the newly
commissioned LHC at CERN joining soon) because it is widely
believed that in such collisions a new form of hadronic matter,
the so called {\it Quark-Gluon-Plasma} (QCD),
will be produced  \cite{QGP}.\\

So far all hydrodynamic models in this field were based on the
usual Boltzmann-Gibbs (BG) form of statistical mechanics. The only
works discussing some general features of the nonextensive
hydrodynamics, which we are aware of \cite{q-hydro}, use a
nonrelativistic approach and are therefore not suitable for the
applications we are interested in. On the other hand it is known
that an approach based on non-extensive statistical mechanics
(used mainly in the form proposed by Tsallis \cite{Tsallis} with
only one new parameter, the nonextensivity parameter $q$)
describes different sets of data in a better way than the usual
statistical models based on BG statistics, cf., \cite{Tsallis} for
general examples and
\cite{q_interpretation,q_interpretation_1,q_approaches,q_approaches_others,NUWW,C_V_Q}
for applications to multiparticle production processes. Roughly
speaking, all observed effects amount to a broadening of the
respective spectra of the observed secondaries (both in transverse
momentum space and in the rapidity space), they take the form of
$q$-exponents instead of the naively expected usual exponents:
$\exp(-X/T) \Longrightarrow \exp_q(-X/T) =
[1-(1-q)X/T]^{1/(1-q)}$. From these studies emerged a commonly
accepted interpretation of the nonextensivity parameter $q$ (in
fact, $|q-1|$) as the measure of some intrinsic fluctuations
characteristic for the hadronizing systems under consideration
\cite{q_interpretation,q_interpretation_1}. For $q > 1$ and in the
transverse momentum space it could be fluctuation of the
temperature $T$ corresponding to some specific heat parameter $C$.
In this case $q - 1 = C$ and therefore it should be inversely
proportional to the volume of the interaction region. This effect
is indeed observed \cite{C_V_Q}. In rapidity space these are
fluctuations of the so called {\it partition temperature}, $T_{pt}
= E/\langle n\rangle$ \cite{NUWW}, which are precisely the same
fluctuations that lead to the Negative Binomial form of the
observed multiplicity distributions, $P(n;\langle n\rangle;k)$,
with its characteristic parameter $k$ given by $k = 1/(q-1)$
\cite{NUWW,Q}. In the case of $q<1$, the interpretation is not at
present clear \cite{q_interpretation_1}. It seems that the first
role of the parameter $q$ is to restrict the allowed phase space
\cite{Cosmic}. Actually, the conjecture associating $q$ with
fluctuations has already been formalized as a new branch of
statistical mechanics called {\it superstatistics} \cite{SuperS}.
It should be noted that there are also arguments connecting
nonextensivity with some special dynamical correlations existing
in the system under consideration \cite{Correlations}, but their
connection with fluctuations is not yet fully clear (cf.,
\cite{Q})\endnote{One should keep in mind, however, that there are
dualities in the non-extensive approach, i.e., that both $q$ and
$1/q$ can be used as the nonextensivity parameter depending on the
normalization of original or $q$-powered probabilities. Also, when
considering the particle-hole symmetry in the $q$-Fermi
distribution, $f(-E,T,\mu,q) = 1-f(E,T,-\mu,2-q)$, in a plasma
containing both particles and antiparticles both $q$ and $2 - q$
occur ($\mu$ denotes the chemical potential here). These dual
possibilities warn us that a theory requiring that only $q
> 1$ has physical meaning is still incomplete \cite{Tsallis}.
These points deserve further considerations which are, however,
going outside the scope of this paper. On the other hand, in the
present paper we are not considering plasmas containing both
particles and antiparticles but only massive pions which are
assumed to obey the $q$-Boltzmann distribution.}.\\

The physical picture emerging from the above experience is that,
instead of a strict local thermal equilibrium customarily assumed
in all applications of statistical models (including
hydrodynamic), one rather encounters {\it a kind of stationary
state, which already includes some interactions}. It can be
introduced in different ways. For example, in \cite{SR} it was a
random distortion of the energy and momentum conservation caused
by the surrounding system. This results in the emergence of some
nonextensive equilibrium. In \cite{Biro,BiroK} the two-body energy
composition is replaced  by some generalized energy sum
$h(E_1,E_2)$, which is assumed to be associative but which is not
necessarily simple addition and contains contributions stemming
from pair interaction (in the simplest case). It turns out that,
under quite general assumptions about the function $h$, the
division of the total energy among free particles can be done.
Different forms of the function $h$ lead then to different forms
of the entropy formula, among which one encounters also the known
Tsallis entropy. The origin of this kind of thinking can be traced
back to the analysis of the $q$-Hagedorn model proposed
some time ago in \cite{Beck}.\\

All phenomenological applications of hydrodynamic models to the
recent multiparticle production processes show that, although
perfect (nonviscous) hydrodynamics successfully describes most
RHIC data \cite{recenthydro,phydroRHIC}, there are indications
that a hadronic fluid cannot be totally ideal. For example, the
perfect fluid dynamical calculations with color glass condensate
initial state could not reproduce the elliptic flow data
\cite{nonviscousindic} indicating a necessity to use some kind of
viscous fluid description. A nonrelativistic viscous fluid is
usually described by the first order Navier-Stokes equations.
However, one needs their special relativistic generalization and
this turns out to be acausal and unstable \cite{BV} (see also
\cite{Tsumura2007,HiscockPhysRevD31,KoidePRC75}). One therefore
looks towards the extended, second order theories accepting all
problems connected with their formulation and practical
applications \cite{Eckart1940,IsraelAnnPhys118,MurongaPRC69,
BaierPhysRevC73,ChaudhuriPRC74, HeinzPRC73,DumitruPRC76}.
Physically, the difference is in that first order theories are
based on the local equilibrium hypothesis, in which the
independent variables are used, whereas in higher order theories
the fluxes of the local equilibrium theory appear as independent
variables. In particular, the entropy vector is quadratic in the
fluxes, containing terms characterizing the deviation from local
equilibrium. This situation plus our experience with the
nonextensive formalism
\cite{q_interpretation,q_interpretation_1,q_approaches,NUWW,C_V_Q,Q}
prompted us to investigate the simple nonextensive formulation of
the perfect hydrodynamic model, a perfect $q$-hydrodynamics
\cite{Osada2007}. It turned out that this describes the
experimental data fairly well. In addition, an apparently
unexpected feature appeared, namely the possibility that
(relatively simple and first order) perfect $q$-hydrodynamics can
serve as a model of (second order and complicated in practical
use) viscous $d$-hydrodynamics. This is the point we would like to
discuss in more detail in this work.\\

In the next Section we shall, for completeness, present the main
points of $q$-hydrodynamics \cite{Osada2007}. In Section 3 we
propose a nonextensive/dissipative conjecture (NexDC), which
allows us to connect ideal $q$-hydrodynamics with a
$d$-hydrodynamics. Consequences of NexDC are discussed in Section
4 (entropy production) and in Section 5 (transport coefficients).
Section 6 contains the summary.

\section{Basic elements of $q$-hydrodynamics}

As in \cite{Osada2007}, we shall limit ourselves to a $1+1$
dimensional baryon-free version of hydrodynamic flow. This is
derived following Lavagno \cite{LavagnoPhysLettA301}, in which a
nonextensive version of the Boltzmann equation has been proposed
and investigated. Because no external currents are assumed, this
corresponds to a kind of {\it perfect $q$-fluid}. There are two
important points in \cite{LavagnoPhysLettA301}: $(i)$ the
Boltzmann equation is formulated for $f_q^q(x,p)$ distribution
rather than for the usual $f_q(x,p)$; $(ii)$ the usual molecular
chaos hypothesis is now assumed in nonextensive form:
\begin{eqnarray}
  h_q[f_q,f_{q1}] = \exp_q \left[ \ln_q f_q + \ln_q f_{q1}\right]
  \quad{\rm where}\quad \exp_q(X) = [1 + (1-q)X]^{\frac{1}{1-q}},\quad \ln_q(X)
=\frac{X^{(1-q)}-1}{1-q} .\label{q-molec-chaos}
\end{eqnarray}
Here $f_q(x,p)$ is the $q$-version of the phase space distribution
function, whereas $h_q[f_q,f_{q1}]$ is the $q$-version of the
correlation function related to the presence of two particles in
the same space-time position $x$ but with different four-momenta
$p$ and $p_1$, respectively. By postulating
Eq.~(\ref{q-molec-chaos}) we are, in fact, {\it assuming} that,
instead of a strict (local) equilibrium, {\it a kind of stationary
state is being formed}, which already includes some interactions
and which is summarily characterized by a parameter $q$; very much
in the spirit of \cite{SR,Biro,BiroK,Beck} mentioned before. With
this assumption we are already departing from the picture of the
usual ideal fluid with its local thermal equilibrium, which is the
prerogative of ideal hydrodynamics \cite{BV}. The consequences of
this fact will be discussed below. The most important ingredient
for further discussion is now the corresponding nonextensive
entropy ($q$-entropy) current
\cite{Osada2007,LavagnoPhysLettA301}:
\begin{eqnarray}
 \sigma_q^{\mu}(x)  =
-k_{\rm B} \!\int \!\! \frac{d^3p}{(2\pi \hbar )^3}
\frac{p^{\mu}}{p^0} \Big\{ f_q^q \ln_q f_q  -f_q  \Big\}.
\label{eq:q_entropy}
\end{eqnarray}
It turns out that $\partial_{\mu}\sigma_q^{\mu} \ge 0$ at any
space-time point, i.e., the relativistic local $H$-theorem is
valid in this case
\cite{LimaPhysRevLett86,LavagnoPhysLettA301,Osada2007}
\endnote{Actually, this form of $q$-entropy current differs
slightly from that in \cite{LavagnoPhysLettA301}. The reason is
that only with such form we can, at the same time, both satisfy
the $H$-theorem and reproduce thermodynamical relations resulting
in our $q$-enthalpy, cf. Eq. (\ref{eq:enthalpy_correspondence}),
((cf. also Eqs. (17a) and (17b) in \cite{Osada2007}), which are
crucial in all further derivations and which were not addressed in
\cite{LavagnoPhysLettA301}).} Assuming $\partial_{\mu}
\sigma_q^{\mu}\equiv 0$ one finds that ($k_B$ is Boltzmann
constant):
\begin{eqnarray}
  f_q(x,p) \!\!&=&\!\! \left[ 1-(1-q)\frac{p_{\mu}u_{q}^{\mu}(x)}{k_{\rm B}
  T_q(x)}\right]^{1/(1-q)} \equiv \exp_q\left[{-\frac{p_{\mu}u_q^{\mu}(x)}{k_B
    T_q(x)}}\right],
\label{eq:q-equilibrium}
\end{eqnarray}
where $T_q(x)$ is the $q$-temperature \cite{Osada2007} and
$u_{q}^{\mu}(x)$ is the $q$-hydrodynamic flow four-vector.
Actually, one should be aware of the fact that there is still an
ongoing discussion on the meaning of temperature in nonextensive
systems. However, the small values of the parameter $q - 1$
deduced from data allow us to argue that, in the first
approximation, $T_q$ can be regarded as the hadronizing
temperature in such a system (cf., \cite{T_q} for a thorough
discussion of the temperature of nonextensive systems). Finally,
we get the $q$-version of the local energy-momentum conservation
\cite{LavagnoPhysLettA301},
\begin{eqnarray}
      \partial_{\nu} {\cal T}_{q}^{\mu\nu}(x) =0 ,\quad {\rm
      with}\quad
   {\cal T}_{q}^{\mu\nu}(x) \equiv \frac{1}{(2\pi\hbar)^3}
   \!\!\int\!\!\frac{d^3p}{p^0} ~p^{\mu}  p^{\nu} f_q^q(x,p)\quad
   {\rm and}\quad  f_q(x,p)\equiv \exp_q\left[{-\frac{p_{\mu}u_q^{\mu}(x)}{k_B
    T_q(x)}}\right].
 \label{eq:Tq}
\end{eqnarray}
In what follows we shall use covariant derivative notation in
which the vector $u^{\mu}$ and tensor $g^{\mu\nu}$ are defined as
\begin{eqnarray}
u^{\nu}_{;\mu} &=& \partial_{\mu} u^{\nu}
  + \Gamma^{\nu}_{\lambda\mu}u^{\lambda}\qquad {\rm  and}\qquad
  g^{\mu\nu}_{;\mu}= \partial_{\mu}g^{\mu\nu}
  +\Gamma^{\mu}_{\sigma\mu}g^{\sigma\nu}
  +\Gamma^{\nu}_{\sigma\mu}g^{\mu\sigma}
\end{eqnarray}
by means of  the Christoffel symbols, $
\Gamma^{\nu}_{\lambda\mu}\equiv \frac{1}{2} g^{\nu\sigma}
(\partial_{\mu}g _{\sigma\lambda} +\partial_{\lambda}g_{\sigma\mu}
-\partial_{\sigma}g_{\lambda\mu})$. In this notation Eq.
(\ref{eq:Tq}) reads:
 \begin{eqnarray}
 {\cal T}_{q;\mu}^{\mu\nu}\, =\,
  \left[({\varepsilon}_q+{P}_q)u^{\mu}_q u^{\nu}_q
  -{P}_qg^{\mu\nu}\right]_{;\mu}\,\, =\,\, \Big[ \varepsilon_q (T_q)
u_q^{\mu}u_q^{\nu} - P_q (T_q)\Delta_q^{\mu\nu} \Big]_{;\mu}\, =\,
0, \label{eq:important}
\end{eqnarray}
where $\Delta_q^{\mu\nu}\equiv g^{\mu\nu}- u_q^{\mu}u_q^{\nu}$.
Here it was assumed that the $q$-modified energy-momentum tensor
${\cal T}_q^{\mu\nu}$ can be decomposed in the usual way in terms
of the $q$-modified energy density and pressure, $\varepsilon_q
\equiv u_{q\mu} {\cal T}_q^{\mu\nu} u_{q\nu}$ and $P_q \equiv
-\frac{1}{3} {\cal T}_{q}^{\mu \nu} \Delta_{q\mu \nu} $, by using
the $q$-modified flow $u_q^{\mu}$ (such that in the rest frame of
the fluid $u_q^{\mu}=(1,0,0,0)$ and for $q\rightarrow 1$ it
becomes the usual hydrodynamic flow $u^{\mu}$). Notice that Eq.
(\ref{eq:important}) is {\it formally} identical to the perfect
hydrodynamic equation but with all the usual ingredients replaced
by their $q$-counterparts (perfect means here that there is
nothing on the r.h.s. of Eq.~(\ref{eq:important})). In this sense
we can
speak about the {\it perfect $q$-fluid} mentioned before.\\

Some remarks are in order before proceeding further.
\begin{itemize}

\item When applied to multiparticle  production processes, each
hydrodynamic model is supplemented with three ingredients, which
must be considered together with Eq. (\ref{eq:important}) (or its
equivalent): $(i)$ - initial conditions (IC) setting the initial
energy density which is going to evolve hydrodynamically; $(ii)$ -
equation of state (EoS) reflecting the internal dynamics of the
fluid considered and $(iii)$ freeze-out conditions describing
transformation of the expanded fluid into observed hadrons. The
problem is that all of them can, in principle, enter with their
own intrinsic fluctuation pattern, i.e., with their own parameters
$q$. In \cite{Osada2007} where we provided preliminary comparison
with experimental data, we have assumed, for simplicity,  the same
value of parameter $q$ throughout the whole collision. It is still
to be checked how good this assumption is. However, in our present
discussion this point is unimportant.

\item Whereas in the usual perfect hydrodynamics (based on the BG
statistics) entropy is conserved in the hydrodynamic evolution,
both locally and globally, in the nonextensive approach it is only
conserved locally. The total entropy of the whole expanding system
is not conserved, because for any two volumes of the fluid,
$V_{1,2}$, one finds that $ S_{q}^{(V_1)} + S_{q}^{(V_2)} \ne
S_{q}^{(V_1\oplus V_2)}$ (where $S_{q}^{(V)}$ are the
corresponding total entropies). This should be always remembered
(albeit, strictly speaking, the hydrodynamic model requires only
local, not global, entropy conservation). As a consequence of this
fact, as we shall see below, contrary to the situation in usual
perfect hydrodynamics, in the perfect $q$-hydrodynamics the
entropy is produced (but not $q$-entropy).

\item To guarantee that hydrodynamics makes sense, there should
exists some spacial scale $L$ such that the volume $L^3$ contains
enough particles. However, in the case when there are fluctuations
and/or correlations characterized by some typical correlation
length $l$ for which we expect that $l> L$, one has to use
nonextensive entropy $S_q^{(L^3)}$ and its (locally defined)
density, $s_q(x) = S_q^{(L^3)}/L^3$. When formulating the
corresponding $q$-hydrodynamics one takes the limit $L\to 0$, in
which case the explicit dependence on the scale $L$ vanishes,
whereas the correlation length $l$ leaves its imprint as a
parameter $q$. In this sense, perfect $q$-hydrodynamics can be
considered as preserving causality and nonextensivity $q$ is then
related with the correlation length $l$ (one can argue that, very
roughly, $q \sim l/L_{\rm eff}  \geq 1$, where $L_{\rm eff} $ is
some effective spacial scale of the $q$-hydrodynamics). If the
correlation length $l$ is compatible with the scale $L_{\rm eff}$,
i.e., $l \approx L_{\rm eff}$, one recovers the condition of the
usual local thermal equilibrium and in this case the
$q$-hydrodynamics reduces to the usual (BG) hydrodynamics. The
above considerations were limited to the case of $q > 1$ only, for
which, as was said before, the clear correspondence with
fluctuations was found. In what concerns the case of $q <1 $, it
seems that following \cite{Cosmic}, where $q < 1$ was found as a
main factor closing the allowed phase space, we can at the moment
only propose that it could probably correspond to the case where,
for some reason, the scale $L$ cannot vanish but must stop at some
value $l < L$. In this case, analogously to what was said above,
one could expect that, again, $q \sim l/L_{\rm eff}$, which this
time would be smaller than unity. We shall not discuss this
possibility further in this work.
\end{itemize}

\section{Nonextensive/dissipative conjecture (NexDC)}

As seen in Eq. (\ref{eq:important}), the structure of the perfect
$q$-hydrodynamical flow is formally identical with the flow of
ideal fluid described by the usual ideal hydrodynamics. We
therefore regard the fluid described by Eq. (\ref{eq:important})
as a {\it perfect $q$-fluid} (in the sense already mentioned
before). Our experience with applications of $q$ statistics
\cite{q_approaches,q_approaches_others} tells us that in cases of
interest to us $|q-1| << 1$. It is therefore tempting to simply
expand the corresponding quantities in powers of $|q-1|$ and to
only keep the linear term \cite{Beck}. The result one gets looks
promising, namely
\begin{eqnarray}
{\cal T}_{q}^{\mu\nu} &\equiv& {\cal T}_{q=1}^{\mu\nu} +(q-1)
{\tau}_q^{\mu\nu}\quad {\rm where} \label{eq:expansion} \\
{\cal T}_{q=1}^{\mu\nu} &\equiv& \frac{g}{(2\pi)^3}\int
\frac{d^3p}{p^0} ~p^{\mu}  p^{\nu} \exp\left(-\frac{p\cdot
u}{T}\right)\quad {\rm and}\quad \tau_q^{\mu\nu} \equiv
\frac{1}{2} \frac{g}{(2\pi)^3} \int
\frac{d^3p}{p^0}~p^{\mu}p^{\nu} \left(\frac{p\cdot u}{T}\right)^2
\exp\left(-\frac{p\cdot u}{T}\right), \label{eq:expansion_1}
\end{eqnarray}
with ${\cal T}_{q=1}$ being the usual energy-momentum tensor
describing an ideal fluid in the BG approach and $\tau_q^{\mu\nu}$
representing a viscous correction caused by the nonextensivity.
However, in order for Eq. (\ref{eq:expansion}) to be valid in the
whole phase space, the $|1-q|\left(\frac{p\cdot u}{T}\right)^2 <
2$ inequality must hold. This means that either such a procedure
can be applied only to a limited domain of phase space, or that $q
= q(x,p)$, a possibility which is outside the scope of the present
work. We must therefore proceed in a more general way. Let us
formally decompose $\varepsilon_q$, $P_q$ and $u^{\mu}_q$ in Eq.
(\ref{eq:important}) into the, respectively, extensive and
nonextensive parts:
\begin{eqnarray}
\varepsilon_q(T_q) \equiv \varepsilon(T_q) +
\Delta\varepsilon_q(T_q), \qquad P_q(T_q) \equiv P(T_q)+\Delta P_q
(T_q), \qquad u^{\mu}_q(x) \equiv u^{\mu}(x)+\delta u^{\mu}_q(x)
.\label{eq:decu}
\end{eqnarray}
Actually this can be only done approximately because our extensive
$\varepsilon$ and $P$ still depend on $T_q$, rather than on $T$.
We tacitly assume that $T_q$ is not too far from $T$ (so far they
are independent parameters but later on we shall impose a
condition on them, see Eq. (\ref{eq:Nex/diss}) below). In Eq.
(\ref{eq:decu}) the four-velocity $u^{\mu}(x)$ is formally a
solution of the equation to which Eq. (\ref{eq:important}) is
transformed by using Eq. (\ref{eq:decu}):
\begin{eqnarray}
\left[\tilde{ \varepsilon}(T_q) u^{\mu}u^{\nu}
 -\tilde{P}(T_q)\Delta^{\mu\nu}
 + 2W^{(\mu}u^{\nu)}
 +\pi^{\mu\nu} \right]_{;\mu} \, =\, 0 .
\label{eq:decomposition}
\end{eqnarray}
One can see that it has form of a dissipative hydrodynamic
equation \cite{Eckart1940,IsraelAnnPhys118,MurongaPRC69,
BaierPhysRevC73,ChaudhuriPRC74, HeinzPRC73,DumitruPRC76} (here
$\Delta^{\mu\nu} \equiv g^{\mu\nu}- u^{\mu}u^{\nu}$ and $A^{(
\mu}B^{\nu )}\stackrel{def}{\equiv} \frac{1}{2}(A^{\mu}B^{\nu} +
A^{\nu}B^{\mu}$)) where $\tilde{\varepsilon}$ is the energy
density, $\tilde{P}$ is the pressure, $W^{\mu}$ denotes the energy
or heat flow vector, whereas $\pi^{\mu\nu}$  is the shear
(symmetric and traceless) pressure tensor. They are defined as
(using the angular bracket notation: $a^{<\mu}b^{\nu
>} \stackrel{def}{\equiv} \left[ \frac{1}{2}(
\Delta^{\mu}_{\lambda} \Delta^{\nu}_{\sigma} +
\Delta^{\mu}_{\sigma} \Delta^{\nu}_{\lambda} )
-\frac{1}{3}\Delta^{\mu\nu} \Delta_{\lambda\sigma} \right]
a^{\lambda}b^{\sigma}$):
\begin{eqnarray}
 \tilde{\varepsilon} = \varepsilon_q+3\Pi,\quad \tilde{P} = P_q  +\Pi,\quad {W}^{\mu}
 = w_q[1+\gamma] \Delta^{\mu}_{\lambda} \delta u_q^{\lambda},\quad
 {\pi}^{\mu\nu\,} = \frac{W^{\mu} W^{\nu}}{ w_q [ 1+\gamma ]^2}
 +\Pi\Delta^{\mu\nu} = w_q~ \delta u_q^{<\mu}  \delta u_q^{\nu
 >},\label{eq:deftilde}
 \end{eqnarray}
and expressed in terms of $\Pi$ (being a kind of a $q$-dependent
bulk pressure), $q$-enthalpy $w_q$ and a $q$-dependent variable
$\gamma$,
 \begin{eqnarray}
 \Pi  \equiv \frac{1}{3} w_q [\gamma^2 + 2\gamma],\qquad\quad
w_q\, \equiv\, \varepsilon_q + P_q \qquad \quad \gamma \equiv\,
u_{\mu}\delta u_q^{\mu} =  -\frac{1}{2}\delta u_{q\mu} \delta
u_q^{\mu}.\label{eq:Piwqgamma}
 \end{eqnarray}
Notice that when $q \rightarrow 1$, the difference between the $q$
and ideal hydrodynamic flows vanishes, $\delta_q^{\mu} \rightarrow
0$, and with it also $\gamma \rightarrow 0$. It means that all
dissipative fluxes of $d$-hydrodynamics (\ref{eq:decomposition})
which are induced by the $q$-flow, like $W^{\mu}$, $\pi^{\mu \nu}$
and $\Pi$, vanish in this limit as well, and one recovers the
equations of the usual perfect hydrodynamics. The variable
$\gamma$ is easier to handle and to calculate than the differences
in flows (see Eq. (\ref{eq:form_of_gamma}) below for its explicit
form). Notice that, whereas the time evolution of $\Pi$ is
controlled by $q$-hydrodynamics (via the respective time
dependencies of $\varepsilon_q$, $P_q$ and $\gamma$) its form is
determined by the assumed constraints which must assure that the
local entropy production is never negative (as is always assumed
in the standard $2^{nd}$ order theory
\cite{IsraelAnnPhys118,MurongaPRC69} ).\\

Now comes the crucial point. To finally link the usual
$q$-hydrodynamics and its $d$ counterpart, one has to fix somehow
the temperature $T_q$ and the flow velocity field $u_q$. This is
done by {\it assuming} that there exists such a temperature $T$
and velocity difference $\delta u_q^{\mu}$ that the following two
relations are satisfied:
\begin{eqnarray}
P(T)\, =\, P_q(T_q),\qquad \qquad \varepsilon(T)\, =\,
\varepsilon_q(T_q) + 3\Pi \label{eq:Nex/diss}.
\end{eqnarray}
We call them  the {\it nonextensive/dissipative} relations, NexDC
in short. Here $\varepsilon$ and $P$ are the energy density and
pressure as defined in the usual BG statistics, i.e., for $q=1$.
Eq. (\ref{eq:Nex/diss}) provides a definite relation between $T_q$
and $T$, therefore, in what follows, we shall mainly use $T$ for a
description of dissipative effects (except of some expressions
when it is easier to keep both $T$ and $T_q$, but always with the
understanding that, because of Eq. (\ref{eq:Nex/diss}) they are
not independent). It is now straightforward to show
\cite{Osada2007} that in this case one can transform Eq.
(\ref{eq:decomposition}) into the equation of the usual
$d$-hydrodynamics:
\begin{eqnarray}
\left\{ \varepsilon(T) u^{\mu}u^{\nu}\, -\, \left[ P(T)+\Pi
\right] \Delta^{\mu\nu}\, +\,  2 W^{(\mu} u^{\nu )}\, +\,
\pi^{\mu\nu} \right\}_{;\mu} \, =0. \label{eq:Nex/diss_equation}
\end{eqnarray}

This completes a demonstration of {\it equivalence} of the perfect
$q$-hydrodynamics represented by Eq. (\ref{eq:important}) and its
$d$-hydrodynamical counterpart represented by Eq.
(\ref{eq:Nex/diss_equation}). It means therefore that the perfect
$q$-fluid is nothing but a viscous fluid which
satisfies the $d$-hydrodynamic equation Eq. (\ref{eq:Nex/diss_equation}).\\

With the bulk pressure $\Pi$ given by Eq. (\ref{eq:Piwqgamma}) and
using the NexDC relations (\ref{eq:Nex/diss}) one can express the
$q$-enthalpy $w_q$ by the usual enthalpy, $w\equiv Ts =
\varepsilon+P$,  and the $q$-dependent variable $\gamma$:
\begin{eqnarray}
 w_q = \varepsilon_q(T_q)\, +\, P_q(T_q)\,  =\, \frac{\varepsilon(T)\, +\, P(T)}
 {[1\, +\, \gamma]^2}\, =\, \frac{w}{[1\, +\, \gamma]^2},
 \label{eq:enthalpy_correspondence}
\end{eqnarray}
Notice that Eq. (\ref{eq:Nex/diss}) leads to the following
important relations between the heat flow vector $W^{\mu}$, the
pressure tensor $\pi^{\mu\nu}$ and the bulk pressure $\Pi$:
\begin{eqnarray}
 W^{\mu} W_{\mu} \, =\,  -3\Pi w,\qquad\quad
\pi^{\mu\nu}W_{\nu} \, =\, -2 \Pi W^{\mu},\qquad\quad
\pi_{\mu\nu}\pi^{\mu\nu} \, =\, 6\Pi^2.
\label{eq:tensor_relations}
\end{eqnarray}

As mentioned before, we expect that in all cases of interest to us
$|q-1| << 1$. This means then that we can regard the state
characterized by $f_q(x,p)$ from Eq.~(\ref{eq:q-equilibrium}) as
some stationary state existing {\it near equilibrium}. This {\it
near equilibrium state} is then defined by the correlation
function $h_q$ in Eq.~(\ref{q-molec-chaos}), for which the energy
momentum tensor can be divided into two parts: the energy-momentum
tensor of the usual ideal fluid, ${\cal T}^{\mu\nu}_{q=1} = {\cal
T}^{\mu\nu}_{\rm eq}$, and the $q$-dependent remaining $\delta
{\cal T}^{\mu\nu}$ (the meaning of all components is the same as
in Eqs. (\ref{eq:deftilde}) and (\ref{eq:Piwqgamma}) and $\delta
{\cal T}^{\mu \nu} \rightarrow 0$ when $q \rightarrow 1$):
\begin{eqnarray}
{\cal T}^{\mu\nu}_q \, \equiv\,  (\varepsilon_q + P_q) u^{\mu}_q
u^{\nu}_q-P_q g^{\mu\nu}\,  \stackrel{def}{\equiv}\,  {\cal
T}^{\mu\nu}_{\rm eq} \, +\, \delta {\cal T}^{\mu\nu}\qquad {\rm
where}\qquad \delta {\cal T}^{\mu\nu} \, =\,  -\Pi\Delta^{\mu\nu}
+ 2W^{(\mu}u^{\nu)} \, +\, \pi^{\mu\nu}.
\label{eq:T_decomposition}
\end{eqnarray}
Using now Eq. (\ref{eq:Nex/diss}) we get from Eq.
(\ref{eq:enthalpy_correspondence}) that
\begin{equation}
\gamma \, =\,  \sqrt{1\, +\, \delta\epsilon_q}\,  -1\, \qquad {\rm
where}\qquad \delta\epsilon_q \, \equiv\, \frac{\varepsilon(T) \,
-\, \varepsilon_q(T_q)}{ \varepsilon_q(T_q)\, +\, P_q(T_q) }.
\label{eq:form_of_gamma}
\end{equation}
This relation allows us to connect the velocity field $u_q$ (which
is solution of the $q$-hydrodynamics given by
Eq.~(\ref{eq:important})) with the velocity field $u$ (which is
solution of the dissipative hydrodynamics given by
Eq.~(\ref{eq:Nex/diss_equation})). To make it more transparent,
let us parameterize these velocity fields by using the respective
fluid rapidities $\alpha_q$ and $\alpha$ {\bf (the metric used is
$g^{\mu \nu} = (1, -1/\tau^2)$)}:
\begin{equation}
u_q^{\mu}(x)\, =\, \left[ \cosh\left( \alpha_q\, -\, \eta
\right),\frac{1}{\tau}\sinh\left( \alpha_q\, -\, \eta \right)
\right],\qquad u^{\mu}(x)\, =\, \left[ \cosh\left( \alpha \, -\,
\eta \right),\frac{1}{\tau}\sinh\left( \alpha \, -\, \eta \right)
\right]. \label{eq:rapidities}
\end{equation}
Because $\gamma\, =\, u_{\mu}\delta u_q^{\mu} \, =\,
\cosh(\alpha_q\, -\, \alpha)\, -\, 1$ the connection between $u$
and $u_q$ is given by
\begin{eqnarray}
\cosh(\alpha_q-\alpha) = \sqrt{1+ \delta\epsilon_q}\qquad
\Longrightarrow\qquad  \alpha= \alpha_q- \log\left(  \epsilon_q  +
\sqrt{1+\delta\epsilon_q} \right) \label{rapidity_condition1}
\end{eqnarray}
(we abandon the other solution, $\alpha= \alpha_q+ \log\left(
\epsilon_q  + \sqrt{1+\delta\epsilon_q} \right)$, because it leads
to the reduction of entropy, not to its production, i.e., for it
$[su^{\mu}]_{;\mu} < 0$, for $q > 1$).

\section{Entropy production in $q$-hydrodynamics}

One of the important implications of the NexDC conjecture is that
in the ideal $q$-hydrodynamics one observes entropy production.
Taking the covariant derivative of Eq.~(\ref{eq:T_decomposition})
and multiplying it by $u_{\nu}$ one gets
\begin{eqnarray}
  u_{\nu}{\cal T}^{\mu\nu}_{q;\mu}\, =\,
  T[su_{\mu}]_{;\mu} \, +\, u_{\nu} \delta {\cal T}_{;\mu}^{\mu\nu}\, =\,
  0\quad \Longrightarrow \quad [s u^{\mu} ]_{;\mu} =-\frac{u_{\nu}}{T}\delta {\cal T}^{\mu\nu}_{;\mu} .
 \label{eq:sprod_picture1}
\end{eqnarray}
This means that, although in the ideal $q$-hydrodynamics the
$q$-entropy is {\it conserved}, i.e., $[ s_q u_q^{\mu}]_{;\mu}=0$,
the usual entropy {\it is produced}; the ideal $q$-fluid is
therefore a kind of usual viscous fluid.
\begin{figure}[h]
\begin{center}
\includegraphics[width=7.5cm]{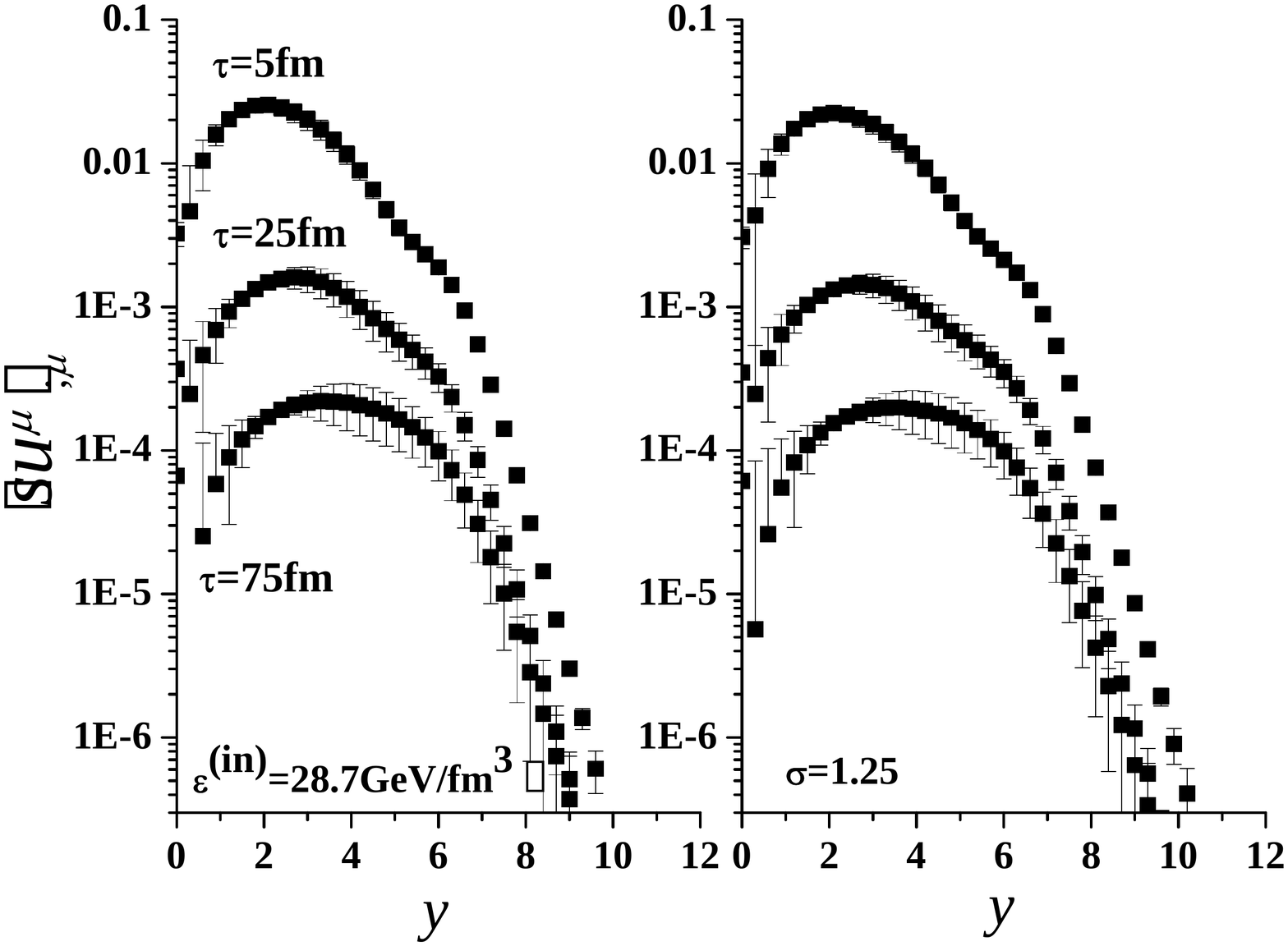}
\includegraphics[width=7.5cm]{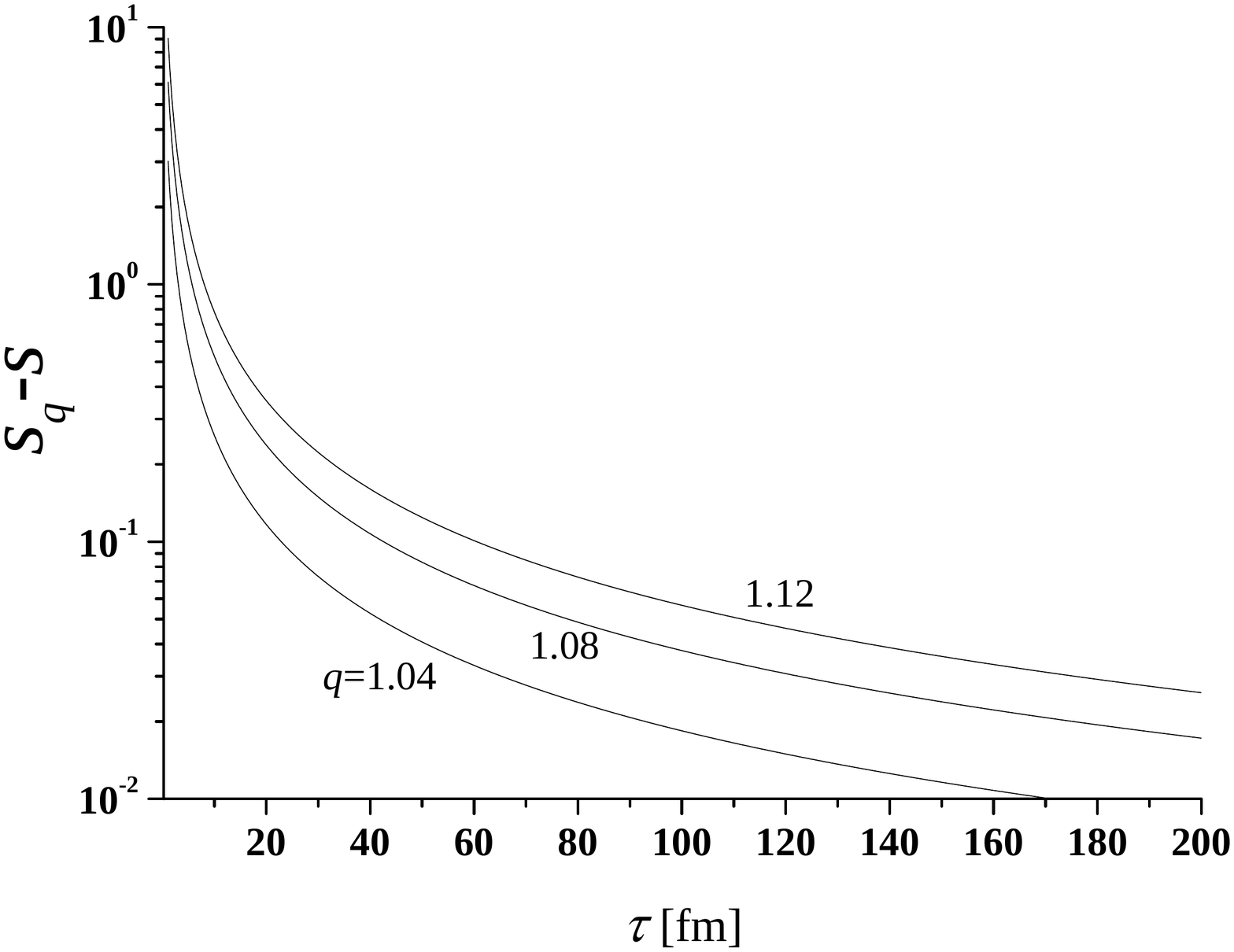}
\vspace{-5mm} \caption{Illustration of different aspects of the
entropy production, $[su^{\mu}]_{;\mu} > 0 $. Two left panels show
production of entropy taking place in different parts of the
longitudinal phase space characterized by the rapidity variable,
$y = (1/2) \ln[(t+z)/(t-z)]$, observed at different evolution
moments $\tau$. First panel shows results for the initial
conditions given by fixed maximal initial energy density
$\varepsilon^{(in)}$, second panel shows results for fixed width
of the initial energy distribution (assumed in gaussian form)
$\sigma$ - see text for other details (cf., also
\cite{Osada2007}). Right (third) panel shows how entropy is
produced in the central region (for $y = 0$, first type of initial
conditions were used here) for different values of the parameter
$q$ and for different evolution moments $\tau$. This entropy
production results in the corresponding growth with $q$ of the
multiplicity of produced secondaries, $N = 11\cdot (7 + 10q)$
(cf., Fig. 8 in \cite{Osada2007}). }\label{Fig1}
\end{center}
\vspace{-0.9cm}
\end{figure}
This is illustrated in Fig.~\ref{Fig1} where the expected entropy
production, as given by Eq.~(\ref{eq:sprod_picture1}), is shown.
All curves presented in Fig. \ref{Fig1} were calculated using the
$q$-hydrodynamical evolution described in \cite{Osada2007} for $q
= 1.08$ with the $q$-dependent initial conditions ($q$-IC). They
were given assuming a $q$-gaussian shape (in rapidity) of the
initial energy density out of which the hydrodynamic evolution
started. Two types of ($q$-IC) were considered: in the first the
maximal initial energy density was fixed as $\varepsilon^{(in)} =
22.3$ GeV/fm$^{3}$ (forcing the width of the $q$-gaussian to be
equal to $\sigma=1.28$), in the second the width of the
$q$-gaussian was fixed as $\sigma = 1.25$ (forcing the initial
energy density to be equal to $\varepsilon^{(in)} = 27.8$
GeV/fm$^{3}$). Because both ($q$-IC) give reasonable fits to
experimental data, they introduce only very small differences in
the entropy production, cf., the left part of Fig. \ref{Fig1}.
These ($q$-IC) were accompanied by the $q$-dependent equation of
state ($q$-EoS) for the relativistic pion gas. Finally, the usual
Cooper-Frye prescription of the freeze-out (but with the
corresponding distribution being the $q$-exponent rather than the
usual one \cite{Osada2007}) was used. All calculations were
performed for $Au+Au$ collisions at $\sqrt{s_{\rm NN}} = 200$ GeV)
(see \cite{Osada2007} where it was shown that $q$-hydrodynamics
with such $q$ and with the respective $q$-IC and $q$-EoS
reproduces $dN/dy$ and $p_{\rm T}$ distributions observed in RHIC
experiment for Au+Au collisions at $\sqrt{s_{\rm NN}}=200$ GeV
energy). Notice that $su^{\mu}_{;\mu}
> 0$ for large $\eta$ region at any $\tau$ (but especially for
the early stage of the hydrodynamical evolution).  It supports
therefore a dissipative character of the $q$-hydrodynamics
mentioned before and leads us to the conclusion that the
equilibrium state generated in heavy-ion collisions may, in fact,
be the $q$-equilibrium state, i.e., some stationary state near the
usual equilibrium state already containing some dissipative
phenomena. Notice that the total multiplicity, which is usually
treated as a measure of entropy, increases with $q$ as expected,
namely $N = 11\cdot (7 + 10q)$.

\section{Transport coefficients of $q$-hydrodynamics.}

The other implication of the NexDC conjecture is that there are
some transport phenomena in an ideal $q$-fluid. We shall now
identify the corresponding transport coefficients and compare them
with the similar coefficients in the usual viscous
$d$-hydrodynamics. Out of a number of different formulations of
$d$-hydrodynamics \cite{Eckart1940,IsraelAnnPhys118,MurongaPRC69,
BaierPhysRevC73,Tsumura2007,HiscockPhysRevD31,ChaudhuriPRC74,
HeinzPRC73,KoidePRC75,DumitruPRC76} we shall choose the $2nd$
order theory of dissipative fluids presented in
\cite{MurongaPRC69,IsraelAnnPhys118}. It does not violate
causality (at least not the global one over a distant scale given
by the relaxation time) and it contains some dissipative fluxes
like heat conductivity and bulk and shear viscosities. We shall
now see, to what extent dissipative fluxes resulting from our
$q$-hydrodynamics can be identified with the heat conductivity and
with the bulk and shear viscosities introduced in
\cite{MurongaPRC69,IsraelAnnPhys118}, and what are the resulting
transport coefficients.\\

Let us start with the most general form of the off-equilibrium
four-entropy current $\sigma^{\mu}$ which in our baryon-free
scenario takes the following form
\cite{MurongaPRC69,IsraelAnnPhys118}:
\begin{eqnarray}
 \sigma^{\mu}\, &=&\, P(T)\beta^{\mu}
 + \beta_{\nu} {\cal T}^{\mu\nu}\, +\,
 Q^{\mu}(\delta {\cal T}^{\mu \nu})\qquad {\rm with}\qquad \beta^{\mu}\, \equiv\,\,
 \frac{u^{\mu}}{T},
   \label{eq:s_Israel}
\end{eqnarray}
where $\delta {\cal T}^{\mu\nu}$ is defined in Eq.
(\ref{eq:T_decomposition}). The function $Q^{\mu} = Q^{\mu} \left(
\delta \cal{T}^{\mu \nu} \right)$ characterizes the
off-equilibrium state, which in our case is induced by the
nonextensive effects and therefore depends on $q$, $Q \rightarrow
0$  when $q \rightarrow 1$. Using the NexDC conjecture, Eqs.
(\ref{eq:Nex/diss}),  and Eq. (\ref{eq:enthalpy_correspondence})
our $q$-entropy current is
\begin{eqnarray}
\sigma_q^{\mu}(\equiv s_q^{\mu})\, =\,  su^{\mu}\,  +\,
\frac{W^{\mu}}{T}\, +\, Q_q^{\mu} \qquad {\rm where}\qquad
Q_q^{\mu} =  Q_{\chi}^{\mu} \equiv \chi \left\{ su^{\mu}\, +\,
\frac{W^{\mu}}{T} \right\} . \label{eq:sigmaq}
\end{eqnarray}
Our {\it near equilibrium state} is thus characterizes by the
function $Q_{\chi}^{\mu}$ in which
\begin{eqnarray}
 \chi \equiv \frac{T}{T_q} \sqrt{1\, -\, \frac{3\Pi}{w}}\, -\, 1\,
 =\, \frac{1}{1 + \gamma}\cdot \frac{T}{T_q} - 1
 \label{eq:Q_expression}
\end{eqnarray}
(temperatures $T$ and $T_q$ are not independent but connected, as
was stated before, by the NexDC conjecture (\ref{eq:Nex/diss})).
We stress that it is given by the dissipative part of our
$q$-system, the one that leads to the increase of the usual
entropy, cf. Fig. \ref{Fig1} (the $q$-entropy, as was discussed
before, remains, however,
strictly conserved, $\sigma^{\mu}_{q;\mu} = 0$). \\

The most general algebraic form of $Q^{\mu}$ for $d$-hydrodynamic
that includes dissipative fluxes up to second order is
\cite{MurongaPRC69}:
\begin{eqnarray}
  Q^{\mu}_{\rm 2nd} \, &=&\,
  \frac{\left[ -\beta_0 \Pi^2\, +\, \beta_1 W_{\nu}W^{\nu}\,
                  -\, \beta_2 \pi_{\nu\lambda}\pi^{\nu\lambda} \right]}{2T} u^{\mu}\,
                  -\, \frac{\alpha_0\Pi W^{\mu}}{T}\, +\,
                  \frac{\alpha_1\pi^{\mu\nu}W_{\nu}}{T},
\end{eqnarray}
where $\beta_{i=0,1,2}$ are the corresponding thermodynamic
coefficients for the, respectively, scalar, vector and tensor
dissipative contributions to the entropy current, whereas
$\alpha_{i=0,1}$ are the corresponding viscous/heat coupling
coefficients. The corresponding expression calculated using the
NexDC conjecture is:
\begin{eqnarray}
 Q^{\mu}_q\, =\,  Q^{\mu}_{\rm 2nd}  \, & = & \,
 \Gamma_{\rm 2nd}\, su^{\mu}\,  +\, \Upsilon_{\rm  1st}
   \frac{W^{\mu}}{T}\\
 {\rm where} && \Gamma_{\rm 2nd}\, \equiv\, -\frac{3\beta_1}{2}\Pi\,
          -\, \frac{(\beta_0+6\beta_2)}{2w} \Pi^2\qquad {\rm and}\qquad
 \Upsilon_{\rm 1st} \, \equiv\,  -\, (\alpha_0\, +\, 2 \alpha_1) \Pi  .\label{eqs:2nd_order_theory}
 \end{eqnarray}
This is given by the second order polynomial in the bulk pressure
$\Pi$. Therefore, it is natural to expect (and this will be our
{\it assumption}) that the most general entropy current in the
NexDC approach is:
\begin{eqnarray}
 Q^{\mu}_{\rm full}\, &=&\, \Gamma(\Pi) su^{\mu}\, +\, \Upsilon(\Pi)
 \frac{W^{\mu}}{T}\, \, =\,\, ( \chi\, +\, \xi )su^{\mu}\, +\, ( \chi\, -\, \xi
 )\frac{W^{\mu}}{T};\label{eq:QGP}\\
 && {\rm where}\qquad \chi \, \equiv\, (\Gamma \, +\, \Upsilon)/2
 \qquad {\rm and}\qquad \xi\ \, \equiv\, (\Gamma\, -\, \Upsilon)/2 \label{eq:CX}
\end{eqnarray}
and where $\Gamma ,\Upsilon$ are (in general infinite) series in
powers of the bulk pressure $\Pi$. In this sense $Q^{\mu}_{\rm
full}$ can be regarded as the {\it full order dissipative current}
in $q$-hydrodynamics. In general one has entropy
production/reduction, i.e., $\sigma^{\mu}_{;\mu} \ne 0$, however,
in the case when $\Gamma(\Pi) = \Upsilon(\Pi)=\chi$ one has
$\sigma^{\mu}_{\chi;\mu} = 0$. Out of the two possible solutions
for $(\Gamma,\Upsilon)$ only one is acceptable,
\begin{equation}
 \Gamma \, =\, 2 \frac{T}{T_q}
\left(\sqrt{1\, -\, \frac{3\Pi}{w}}\, -1\, \right) = -
\frac{2\gamma}{1 + \gamma}\frac{T}{T_q},
    \qquad \qquad \Upsilon \, =\,  2 \frac{T\, -\, T_q}{T_q}, \label{eq:Q_general1}
\end{equation}
because only for it $u_{\mu}Q^{\mu}_{\rm full} \le 0$ (i.e., the
entropy is maximal in the equilibrium \cite{MurongaPRC69}, this is
because $(T - T_q)/T_q$ is always positive for $q\ge 1$
\cite{Osada2007}). In this way we finally arrive at the following
expression for the full order dissipative entropy current emerging
from the NexDC approach \endnote{One must keep in mind therefore
that, although to define the entropy current in $q$-statistics we
require only that $q > 0$,  the NexDC correspondence requires in
addition that $q > 1$ to be consistent with $d$-hydrodynamics
\cite{MurongaPRC69}. Only then, as witnessed by Fig. \ref{Fig1},
constant (nonzero) initial Tsallis entropy results in increasing
BG entropy (demonstrating itself in the total multiplicity
increasing with $(q-1)$). For $q <1$ we could have decreasing BG
entropy from some positive initial value (and, accordingly, total
multiplicity decreasing with $(q-1)$). This point is, however, not
totally clear at present and we plan to address it elsewhere.}:
\begin{eqnarray}
\sigma^{\mu}_{\rm full} \, &\equiv&\,   su^{\mu}\, +\,
\frac{W^{\mu}}{T}\, -\, \frac{2T}{T_q}\left[\, 1\, -\, \sqrt{ 1\,
-\, \frac{3\Pi}{w}}\, \right]su^{\mu}\, +\, \frac{2(T\, -\,
T_q)}{T_q}\frac{W^{\mu}}{T}. \label{eq:full_order_current}
\end{eqnarray}
Limiting ourselves to situations when $T/T_q \approx 1$ and
neglecting terms higher than ${\cal O}(3\Pi/w)^2$, one obtains
that
\begin{eqnarray}
Q^{\mu}_{\rm full}  \, &\approx&\,  \left[ -\left( \frac{3\Pi}{w}
\right)\, -\, \frac{1}{4} \left( \frac{3\Pi}{w} \right)^2 \right]
su^{\mu}. \label{eq:Q_chi_2nd}
\end{eqnarray}
Comparing now Eqs.~(\ref{eqs:2nd_order_theory}) and
(\ref{eq:Q_chi_2nd}) one gets that $ \beta_1\, =\, \frac{2}{w}$, $
\beta_0\, +\, 6\beta_2\, =\, \frac{9}{2w}$ and $\alpha_0\, +\,
2\alpha_1=0$. Since in the Israel-Stewart theory
\cite{IsraelAnnPhys118} the relaxation time $\tau$ is proportional
to the thermodynamic coefficients $\beta_{0,1,2}$, it is naturally
to assume that in the case of NexDC  it is proportional to the
inverse of the enthalpy, $\tau \propto 1/w$ (notice that for the
classical Boltzmann gas of massless particles $\beta_2=3/w$
\cite{MurongaPRC69,DumitruPRC76}).\\

\vspace{-3mm}
\begin{figure}[h]
\begin{center}
\includegraphics[width=12cm]{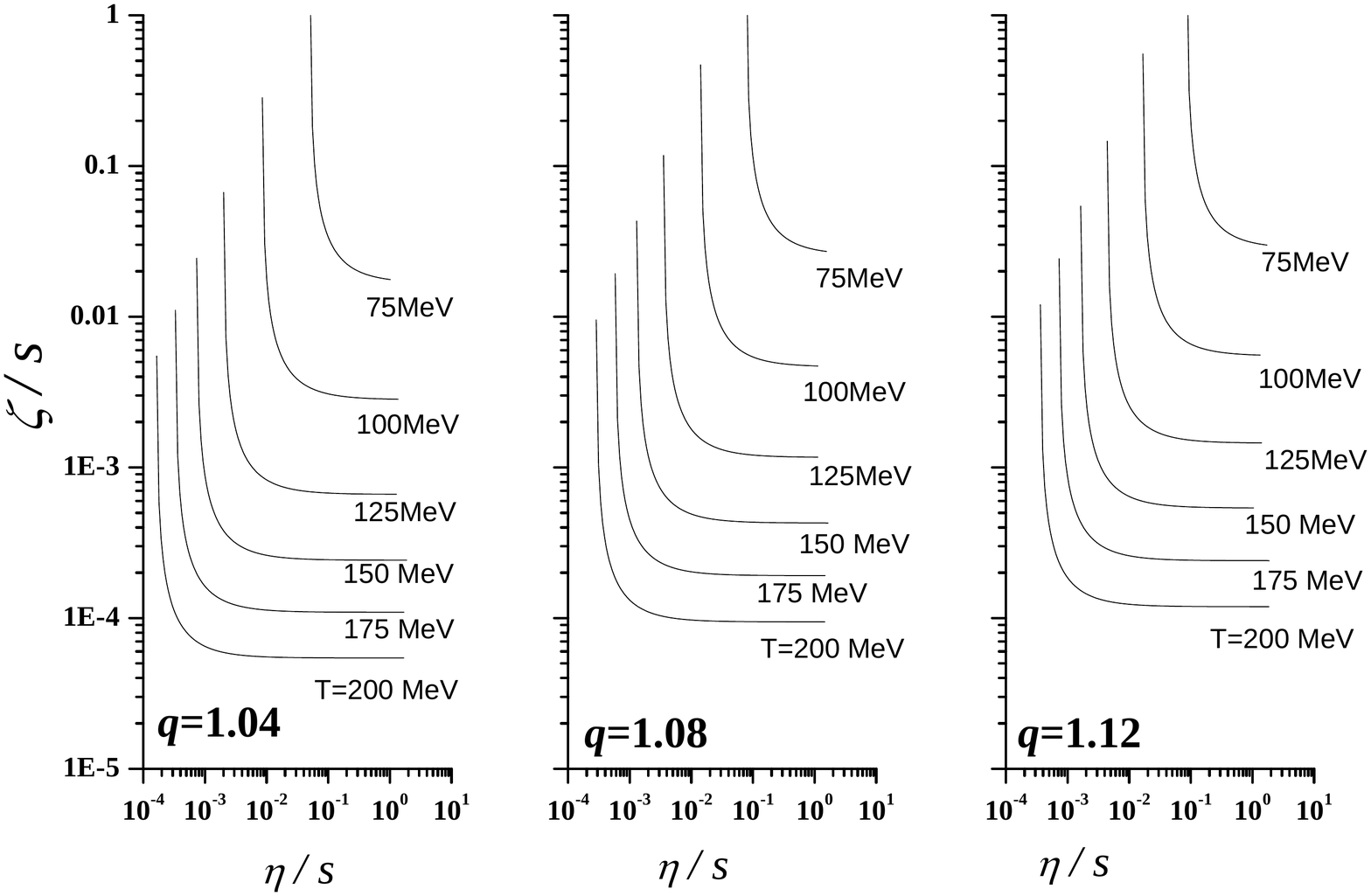}\\
\vspace{-5mm}
\includegraphics[width=7cm]{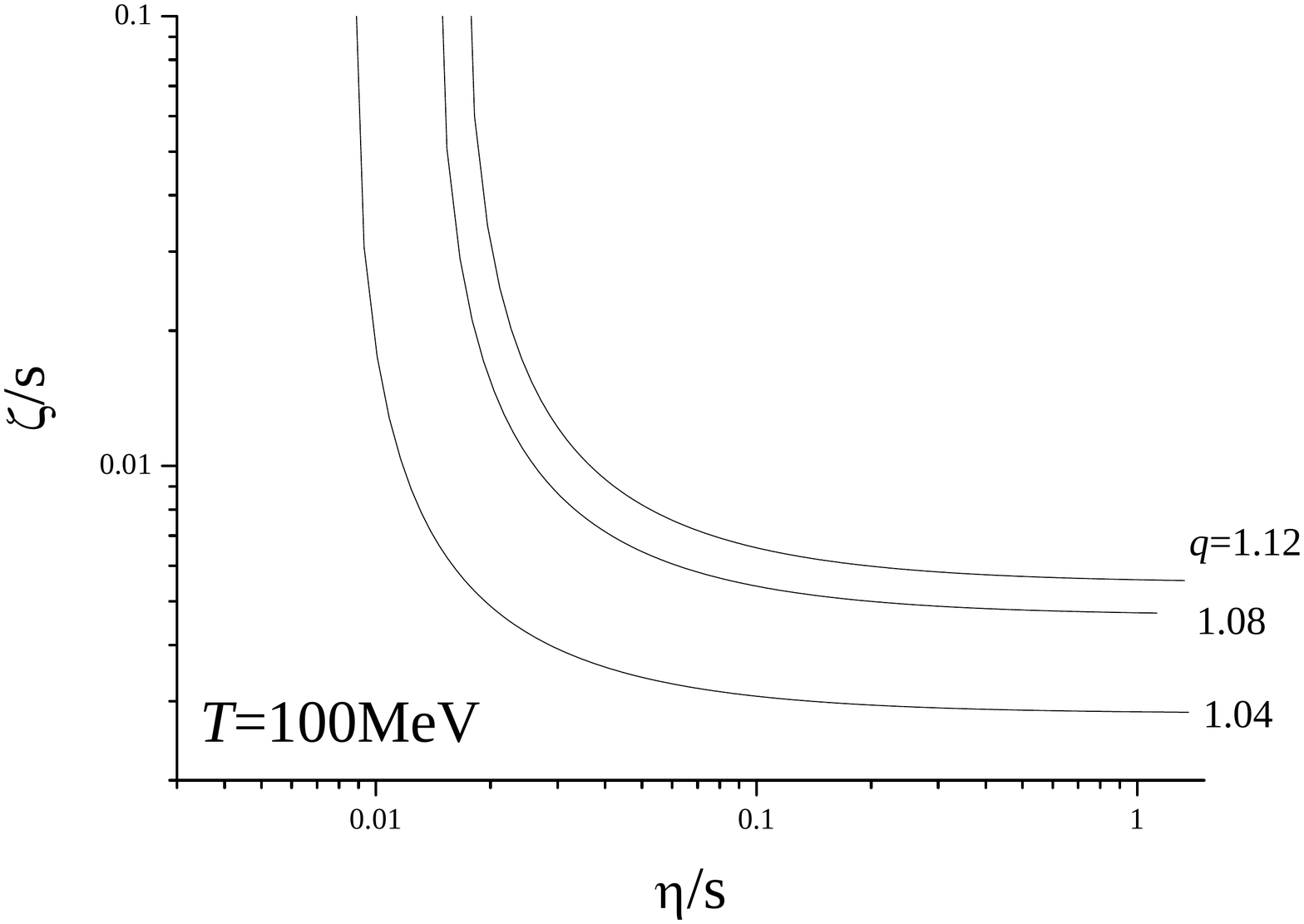}
\vspace{-0.5cm} \caption{Illustration of the sum rule
(\ref{eq:sum_rule}) for different values of temperature $T$ and
different nonextensivities $q$ (the other parameters are the same
as in Fig. \ref{Fig1} for the first type of initial conditions).
}\label{Fig2}
\end{center}
\vspace{-0.5cm}
\end{figure}

Let us now see what kind of bulk and shear viscosities emerge from
the NexDC approach. To this end let us write the full order
entropy current Eq.~(\ref{eq:QGP}) in the following form:
\begin{eqnarray}
\sigma^{\mu}_{{\rm full};\mu} \, &=&\, [(1\, +\,
\chi)\Phi^{\mu}]_{;\mu}\, +\, [\xi\Psi^{\mu}]_{;\mu}\qquad {\rm
where}\qquad \Phi^{\mu}\, =\, su^{\mu}\, +\,
\frac{W^{\mu}}{T}\qquad {\rm and}\qquad \Psi^{\mu}\, =\,
su^{\mu}\, -\, \frac{W^{\mu}}{T}. \label{eq;sigma_full0}
\end{eqnarray}
Because conservation of the $q$-entropy, $\sigma^{\mu}_{q;\mu}=0$,
is equivalent to $[(1+\chi)\Phi^{\mu}]_{;\mu} =0$, therefore,
using Eq.~(\ref{eq:tensor_relations}) one gets that $\Psi^{\mu} =
 - \frac{W^{\nu}W_{\nu}}{3\Pi T} u^{\mu}
+ \frac{ W_{\nu}}{2\Pi T} \pi^{\mu\nu}$, therefore
\begin{eqnarray}
\qquad && \sigma^{\mu}_{{\rm full};\mu} \, = \, -\frac{\Pi}{T}
(wu^{\mu}X_{\mu})\, -\, \frac{W^{\mu}}{T} \tilde{Y}_{\mu}\, +\,
\frac{\pi^{\mu\nu}}{T} Z_{\mu\nu}  \label{eq:sigma_full1}\\
{\rm where}: \qquad && X_{\mu} = -\frac{\xi}{\Pi} \left[
\frac{\partial_{\mu}\Pi}{\Pi} +\frac{\partial_{\mu} T}{T}
+\frac{\partial_{\mu} \xi}{\xi} \right],\qquad\qquad Y_{\mu}
\!\!=\!\! \frac{\xi}{\Pi} \left[ \frac{2}{3} u^{\nu}W_{\mu;\nu} +
\frac{1}{3}W_{\mu}u^{\nu}_{;\nu} -\frac{1}{2}\pi^{\nu}_{\mu;\nu} \right],\\
&& Z_{\mu\nu} = \frac{\xi}{\Pi}\left[  \frac{1}{2} W_{\nu;\mu}
\right],\qquad\qquad\qquad \tilde{Y}_{\mu} =  Y_{\mu} - \Pi
X_{\mu}, \qquad \qquad \tilde{Z}_{\mu\nu}\equiv Z_{\mu\nu}+
\frac{\tilde{Y}_{\mu} W_{\nu}}{2\Pi} .
\end{eqnarray}
Eq.~(\ref{eq:tensor_relations}) allows to eliminate the term
proportional to the heat flow, $\frac{W^{\mu}}{T}$. Finally one
obtains
\begin{eqnarray}
\sigma^{\mu}_{{\rm full};\mu}\, &=&\, -\frac{\Pi}{T}
(wu^{\mu}X_{\mu})\, +\, \frac{\pi^{\mu\nu}}{T}
\tilde{Z}_{\mu\nu}\,\, =\,\, \frac{\Pi^2}{\zeta T}\, +\,
\frac{\pi^{\mu\nu}\pi_{\mu\nu}}{2\eta T} \ge 0 ,
\label{eq:sigma_with_coeff}
\end{eqnarray}
where we have introduced the usual bulk and shear viscosities,
$\zeta$ and $\eta$. Notice that, because of
Eq.~(\ref{eq:tensor_relations}) one avoids the explicit
contribution to the entropy production coming from the heat flow,
$\frac{W^{\mu}}{T}$, which is present in
Eq.~(\ref{eq:sigma_full1}) when one discusses a baryon free fluid,
in which case the necessity to use the Landau frame would appear.
As one can see, Eq.~(\ref{eq;sigma_full0}) is covariant and
therefore it does not depend on the frame used. One arrives at our
main result: the sum rule connecting bulk and shear viscosity
coefficients (expressed as their ratios over the entropy density
$s$), see Fig. \ref{Fig2}:
\begin{eqnarray}
 \frac{1}{\zeta/s}+\frac{3}{\eta/s} = \frac{w \sigma^{\mu}_{{\rm
 full};\mu}}{\Pi^2}.
\label{eq:sum_rule}
\end{eqnarray}
\vspace{-10mm}
\begin{figure}[h]
\begin{center}
\includegraphics[width=7.5cm]{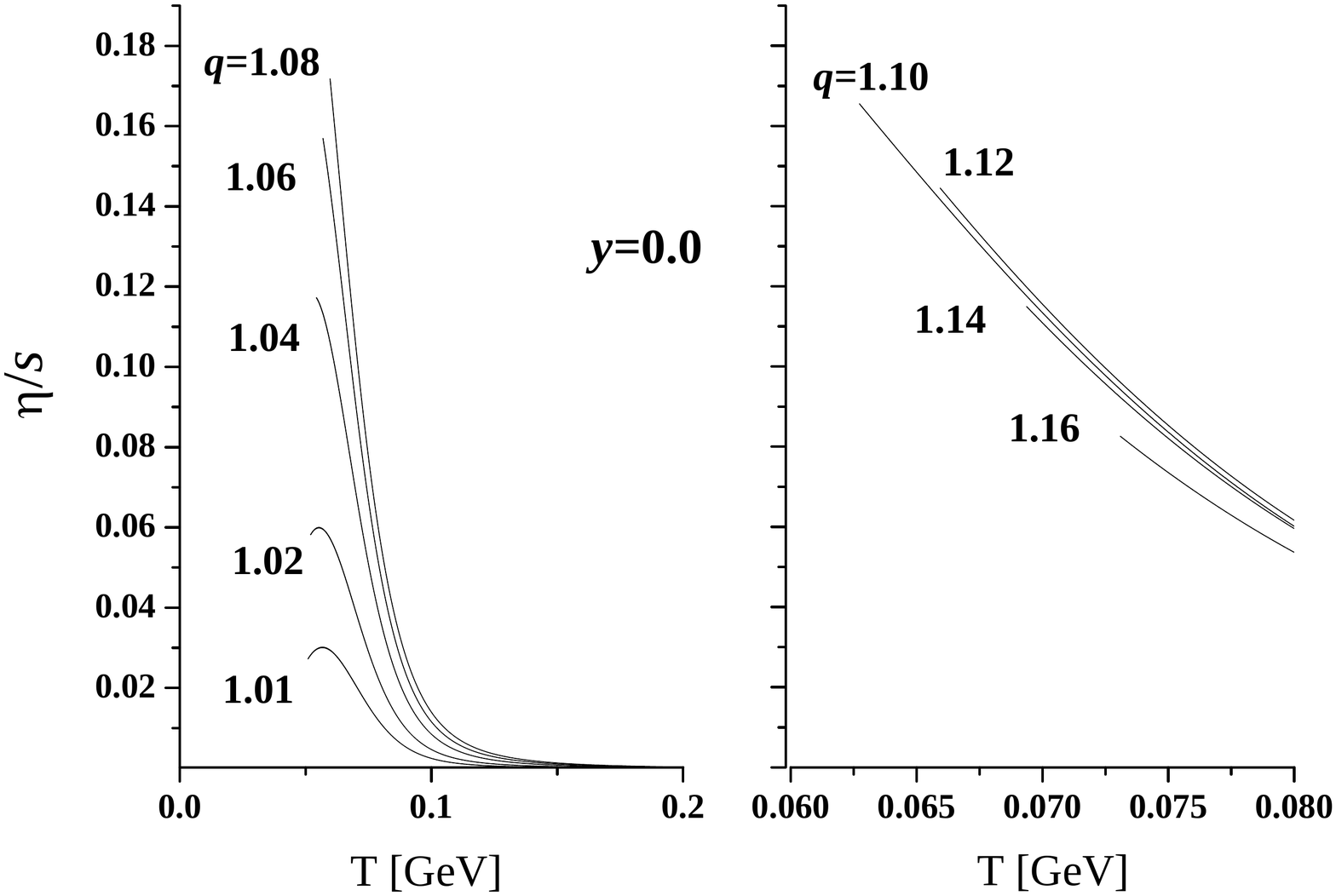}
\includegraphics[width=7.5cm]{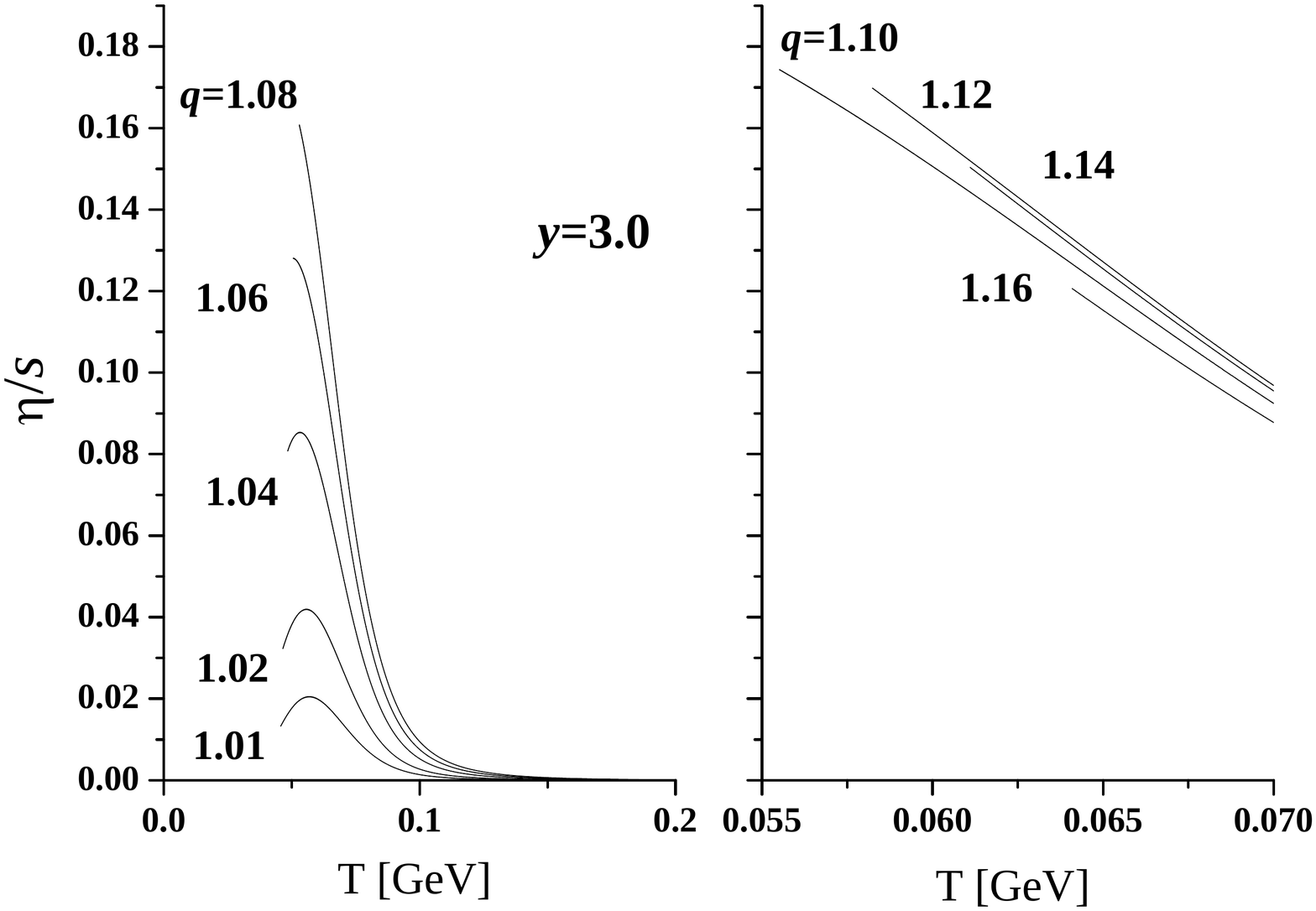}\\
\vspace{-2mm}
\includegraphics[width=8.5cm]{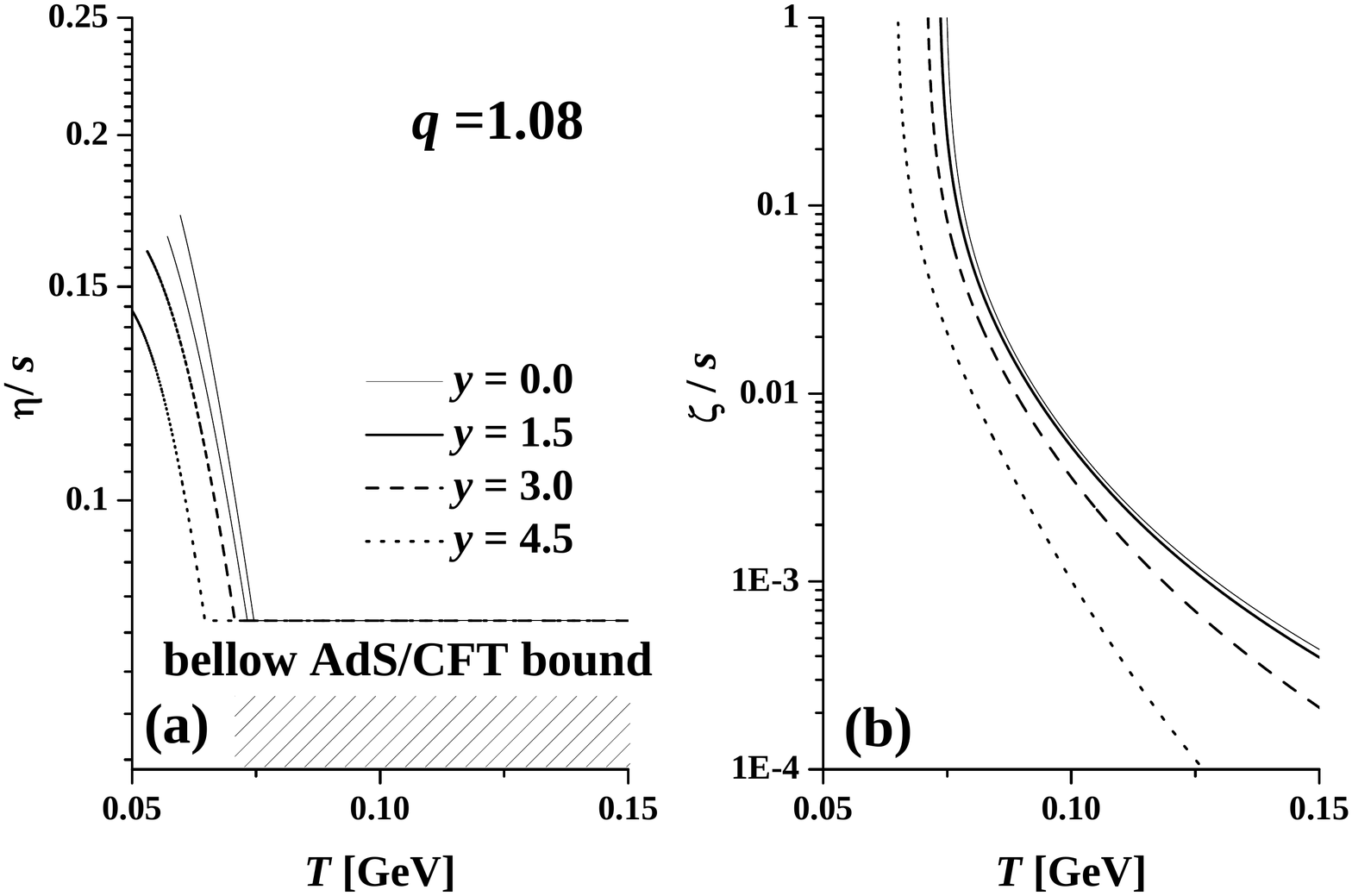}
 \caption{Upper panels: the ratio of the shear  viscosity over the entropy
 density, $\eta/s$, as function of temperature $T$,  calculated using
 Eq. (\ref{eq:nex/diss_prediction}) for different values of $q$ at the
 mid-rapidity region $y=0$ (two upper-left panels) and for rapidity $y = 3$
 (two upper-right panels). The same version of hydrodynamical model was used as in
 Fig. \ref{Fig1} with the first type of initial conditions.
 Notice that for $q \rightarrow 1$ this ratio vanishes,
 as expected, proving therefore correctness of our numerical calculations.
 The lower two panels summarize the rapidity dependence of both the shear
 and the bulk viscosity ratios, $\eta/s$ (left panel) and  $\xi /s$ (right panel).
 The $\eta/s$ is calculated using Eq. (\ref{eq:nex/diss_prediction}), i.e.,
 assuming  $\zeta/s =0$, and presented only above the limit value $\eta/s = 1/4\pi$
 found in the AdS/CFT approach \cite{KovtunPRL94}. The right panel shows
 the bulk viscosity, $\zeta/s$, calculated in this limit,
 i.e., assuming $\eta/s = 1/4\pi$ and using Eq. (\ref{eq:sum_rule}).} \label{Fig3}
\end{center}
\end{figure}

This is as much as one can get from the $q$-hydrodynamics alone.
To disentangle this sum rule, one has to add some additional
input. Suppose, therefore, that we are interested in an extremal
case, when a total entropy is generated by action of the shear
viscosity only. In this case one can rewrite the first part of Eq.
(\ref{eq:sigma_with_coeff}) as
\begin{eqnarray}
 \sigma^{\mu}_{{\rm full};\mu}\, =\, \frac{\pi^{\mu\nu}}{T}
             \left[ - \frac{\pi_{\mu\nu}}{6\Pi}
             (wu^{\lambda}X_{\lambda})\,  +\, \tilde{Z}_{\mu\nu}  \right],
\end{eqnarray}
and arrive at
\begin{eqnarray}
\frac{\eta}{s} = \frac{\gamma (\gamma + 2)}{(\gamma + 1)^2}
\left[ \frac{\pi^{\mu\nu}}{\Pi}\frac{\tilde{Z}_{\mu\nu}}{T} -
su^{\lambda} X_{\lambda} \right]^{-1}.
\label{eq:nex/diss_prediction}
\end{eqnarray}

The predictions of Eq. (\ref{eq:nex/diss_prediction}) are shown in
Fig. \ref{Fig3} for different values of the parameter $q$ and for
different rapidities. They are also confronted with the known
result on $\eta/s$, provided by the popular Ads/CFT conjecture
\cite{KovtunPRL94}, that $\eta/s \ge 1/4\pi$. Assuming the
validity of this limitation we can use Eq.~(\ref{eq:sum_rule})
only in the region where the r.h.s. of
Eq.~(\ref{eq:nex/diss_prediction}) is smaller than (or equal to)
$1/4\pi$ (in which case we put $\eta/s = 1/4\pi$), otherwise
(because of our assumption that the total entropy is generated by
the shear viscosity only) we have to put $\zeta/s=0$ and use
Eq.~(\ref{eq:nex/diss_prediction}) to evaluate $\eta/s$. The
results for $\zeta/s$ and $\eta/s$ are shown in Fig.\ref{Fig3}.
Notice that, when the r.h.s of Eq.~(\ref{eq:nex/diss_prediction})
approaches $1/4\pi$, $\zeta/s$ given by Eq.~(\ref{eq:sum_rule})
approaches infinity. To avoid such a situation, $\eta/s$ should
start to increase at higher temperatures, for example at $T \ge
75$ MeV \endnote{It should be kept in mind that we have so far
obtained only the relations between the Israel-Steward
coefficients ($\beta_{0,1,2}$ and $\alpha_{0,1}$) and not their
individual values, therefore we cannot compare our results to
those of the ideal Boltzmann gas case.  This is due to the tensor
form of relations (\ref{eq:tensor_relations}) and to the fact that
the original perfect $q$-hydrodynamics (Eq. (\ref{eq:Tq})) does
not contain any natural space-time scale. It is then natural that
NexDC conjecture does not introduce {\it per se} any definite
relaxation time or viscous-heat coupling length scale. Also, in
examples shown here, only EoS for pionic gas was used as in
\cite{Osada2007}. Although it was shown there that such EoS
depends only very weakly on the parameter $q$, it remains to be
checked whether this is also true for a more realistic EoS with
quarks and gluons and in the vivinity of the QGP $\rightarrow$
hadronic matter phase transition. We plan to address this point
elsewhere.}.

\section{Summary}

To summarize, we have discussed dissipative hydrodynamics from a
novel point of view. This is provided by the nonextensive
formulation of the usual perfect hydrodynamical model recently
proposed by us \cite{Osada2007}. Such a model can be solved
exactly and contains terms which can be interpreted as due to some
dissipative effects. They can be identified and are expressed by
the nonextensivity parameter $q$ of the Tsallis formalism applied
here. This finding was used to propose a possible full order
expression for the dissipative entropy current $\sigma^{\mu}_{{\rm
full}}$ resulting from the nonextensive approach. The
corresponding bulk and shear transport coefficients resulting from
$q$-hydrodynamics are connected by a kind of sum rule, Eq.
(\ref{eq:sum_rule}). They were calculated for some specific
simplified case, cf. Fig. \ref{Fig3}. We close by noticing that
there is still some uncertainty in the equation of state used in
our numerical example (for example, in what concerns the role of
the possible QGP phase transition, which, when included in EoS,
could considerably affect $\eta/s$ presented in Fig.~\ref{Fig3}).
We plan to address this subject elsewhere.\\

From our point of view it is interesting to observe that (at least
some) remedies proposed to improve the formulation of
$d$-hydrodynamic \cite{KoidePRC75,BV} (like the use of some
induced memory effects) introduce conditions which in statistical
physics lead in a natural way to its nonextensive version
described by the parameter $q$ \cite{Tsallis}. It is then natural
to expect that a nonextensive version of the hydrodynamical model
with the nonextensivity parameter $q$ could help us to circumvent
(at least to some extent) the problems mentioned. It is because
equations of the perfect nonextensive hydrodynamics (or perfect
$q$-hydrodynamics) can be formally solved in an analogous way as
equations of the usual perfect hydrodynamics \cite{Osada2007}.
However, it turns out that, from the point of view of the usual
(extensive) approach the new equations contain terms which can be
{\it formally} identified with terms appearing in the usual
dissipative hydrodynamics ($d$-hydrodynamics). Although this does
not fully solve the problems of $d$-hydrodynamics, nevertheless it
allows us to extend the usual perfect fluid approach (using only
one new parameter $q$) well behind its usual limits, namely toward
the regions reserved so far only for the dissipative approach.


\section*{Acknowledgements}

GW would like to express his gratitude towards the organizers of
the Int. Conf. on Statistical Physics - SigmaPhi2008 - Kolymbari,
Crete, 14-18 July 2008 where this work was presented as an invited
talk. Partial support (GW) from the Ministry of Science and Higher
Education under contract 1P03B02230 is also acknowledged.

\end{document}